\documentclass[twocolumn,showpacs,preprintnumbers]{revtex4}

\usepackage{graphicx}
\usepackage{dcolumn}
\usepackage{bm}

\begin{document}

\title{The Violation of Stokes-Einstein Relation in Supercooled water}
\author{Sow-Hsin Chen\footnote{All correspondence regarding this paper should be addressed to
Sow-Hsin Chen (sowhsin@mit.edu).}$^{\dagger}$, Francesco
Mallamace$^{\dagger \ddagger}$, Chung-Yuan Mou$^{\S}$, Matteo
Broccio$^{\dagger \ddagger}$, Carmelo Corsaro$^{\ddagger}$,
Antonio Faraone$^{\ddagger}$, and Li Liu$^{\dagger}$}

\affiliation{$^{\dagger}$Department of Nuclear Science and
Engineering, Massachusetts Institute of Technology, Cambridge MA
02139 USA} \affiliation{$^{\ddagger}$Dipartimento di Fisica and
CNISM, Universit\`{a} di Messina I-98166, Messina, Italy}
\affiliation{$^{\S}$Department of Chemistry, National Taiwan
University, Taipei, Taiwan}

\date{\today}

\begin{abstract}
By confining water in nanopores, so narrow that the liquid cannot freeze, it
is possible to explore its properties well below its homogeneous nucleation
temperature $\mathbf{T}_{H}\mathbf{\approx 235}$ K. In particular, the
dynamical parameters of water can be measured down to $\mathbf{180}$ K
approaching the suggested glass transition temperature $\mathbf{T}_{g}%
\mathbf{\approx 165}$ K. Here we present experimental evidence, obtained
from Nuclear Magnetic Resonance and Quasi-Elastic Neutron Scattering
spectroscopies, of a well defined decoupling of transport properties (the
self-diffusion coefficient and the average translational relaxation time),
which implies the breakdown of the Stokes-Einstein relation. We further show
that such a non-monotonic decoupling reflects the characteristics of the
recently observed dynamic crossover, at about $\mathbf{225}$ K, between the
two dynamical behaviors known as fragile and strong, which is a consequence
of a change in the hydrogen bond structure of liquid water.
\end{abstract}


\maketitle

Despite its fundamental importance in science and technology, the physical
properties of water are far from being completely understood. The liquid
state of water is unusual especially at low temperatures \cite
{angel,debenedetti,mishima}. For example, contrary to other liquids, water
behaves as if there exists a singular temperature toward which its
thermodynamical properties, such as compressibility, thermal expansion
coefficient, and specific heat, diverge \cite{angel}. The efforts of
scientists from many disciplines to seek a coherent explanation for this
unusual behavior, in combination with its wide range of impacts, make water
one of the most important open questions in science today. On the other
hand, the nature of the glass transition (GT) of water represents another
challenging subject for current research \cite{velikov}. Dynamical
measurements in glass forming liquids have shown a dramatic slowdown of both
macroscopic (viscosity $\eta $ and self-diffusion coefficient $D$) and
microscopic (average translational correlation time $\tau $) observables, as
temperature is lowered towards the GT temperature $T_{g}$. Accordingly, a
comprehension of the GT has been sought through the study of the dynamics at
the molecular level, which, despite all efforts, has not yet been completely
understood \cite{sokolov,goetze,angel1,angell2}. Keeping in mind the
``complexities'' of both low-temperature water and its GT, we present here
direct measurements of two dynamical parameters of water: the self-diffusion
coefficient and the average translational relaxation time, in the
temperature range from $280$ to $190$ K, obtained by Nuclear Magnetic
Resonance (NMR) and Quasi-Elastic Neutron Scattering (QENS) experiments,
respectively.

Bulk water can be supercooled below its melting temperature ($T_{M}$) down
to $\approx 235$ K, below which it inevitably crystallizes; it is just in
such supercooled metastable state that the anomalies in its thermodynamical
properties are most pronounced, showing a power law divergence towards a
singular temperature $T_{S}=228$ K. At ambient pressure, water can exist in
a glassy form below $135$ K. Depending on $T$ and $P$, glassy water has two
amorphous phases with different structures: a low (LDA) and a high (HDA)
density amorphous ice; thus it shows a polymorphism. LDA can be formed from
HDA and vice versa; LDA, if heated, undergoes a glass-to-liquid transition
transforming into a highly viscous fluid, then crystallizes into cubic ice
at $T_{X}$ $\approx 150$ K. Thus an experimentally inaccessible $T$ region
exists in bulk water between $T_{H}$ and $T_{X}$. Experiments performed
within this interval could be of fundamental interest for understanding the
many open questions on the physics of water. For example, the presence of a
first order liquid-liquid transition line (LLTL), the precise location of
its $T_{g}$, recently suggested at about $165$ K \cite{velikov,poole}, and
the existence of a \textit{fragile-to-strong dynamic crossover} (FSC) on
approaching $T_{g}$ from the liquid side \cite{ito}. The existence of a LLTL
leads to conjecture that liquid water possesses a low-temperature second
critical point (predicted to be located at $T_{c}\approx 220$ K$,P_{c}%
\approx 1$ Kbar) \cite{debenedetti}, below which it can switch from one
phase, a high density liquid (HDL), to another phase, a low density liquid
(LDL), whose corresponding vitreous forms are the HDA and LDA, respectively.
The difference between the two liquid phases lies in the water structure: in
the HDL, the local tetrahedrally coordinated hydrogen-bond network is not
fully developed, whereas in the LDL a more open, locally ice-like,
hydrogen-bond network is fully developed \cite{soper}. Thus, near $T_{c}$,
water is a mixture of both LDL and HDL phases associated with a diverging
density fluctuation. At higher temperatures, the two liquid phases are
indistinguishable. Lowering temperature or increasing pressure will result
in an increase of the LDL phase with respect to the HDL phase. The FSC can
be identified by the temperature at which transport properties, like the
shear viscosity $\eta $ or the inverse self-diffusion coefficient $1/D$,
cross over from a non-Arrhenius (fragile) to an Arrhenius (strong) behavior
on approaching $T_{g}$.

A possibility to enter this inaccessible temperature range of water, named
``no-man's-land", is now shown by confining water in nano-size pores \cite
{bergman,tanaka,dore,dore2000}. When contained within these pores, water
does not crystallize, and can be supercooled well below $T_{H}$. Vycor pores
\cite{dore,dore2000}(a porous hydrophilic silica glass), micellar systems or
layered vermiculite clay \cite{bergman} are examples of confining
nano-structures. The latter systems have been used to explore the Arrhenius
behavior of the dielectric relaxation time ($\tau _{D}$) of very deeply
supercooled water.

The FSC was recently confirmed by a QENS experiment, which measured the $T$
and $P$ dependences of the average translational relaxation time $%
\left\langle \tau _{T}\right\rangle $ for water confined in nanopores of
silica glass \cite{chen2004,chen2005}. In particular, as the temperature is
lowered, a $\left\langle \tau _{T}\right\rangle $ versus $1/T$ plot exhibits
a cusp-like crossover from a non-Arrhenius to an Arrhenius behavior at a
temperature $T_{L}(P)$. This crossover temperature decreases steadily upon
increasing $P$, until it intersects the $T_{H}$ line of bulk water at $P\sim
1.6$ Kbar. Beyond this point, the FSC can no longer be identified. These
results, suggestive of the existence of the two liquid phases, have been
explained in a molecular dynamics (MD) simulation study by considering the
existence of a critical point. The MD study shows that the FSC line
coincides with the line of specific heat maxima $C_{p}^{\max }$, called
Widom line. The Widom line is the critical isochore above the critical point
in the one-phase region \cite{stanley}. Moreover, it is observed that
crossing this line corresponds to a change in the $T$ dependence of the
dynamics. More precisely, the calculated water diffusion coefficient, $D(T)$%
, changes according to a FSC, while the structural and thermodynamic
properties change from those of HDL to those of LDL.

\section*{Results and Discussion}

In this report, we present a detailed study done using two
different experimental techniques, neutron scattering and NMR, to
probe dynamical properties of confined water at low temperatures,
well inside the inaccessible region of bulk water. Our main aim is
to clarify the properties of water as a glass forming material,
measuring directly, with NMR spectroscopy, the self diffusion
coefficient $D$ as a function of temperature, and comparing the
obtained results with the translational relaxation time
$\left\langle \tau _{T}\right\rangle $ measured by QENS.
$\left\langle \tau _{T}\right\rangle $ is a quantity proportional
to the viscosity $\eta $. These new measurements enable one to
compare the proportionality of the transport coefficient $1/D$ and
viscosity $\eta $, and can provide a test for the theoretical
description of dynamics at the molecular scale of glass forming
materials. According to Ito et al. \cite{ito}, the FSC can be
intimately connected to the presence of a thermodynamic event in
liquid water: i.e. the temperature dependence of the inverse
self-diffusion coefficient does not follow that of viscosity or
inverse mutual diffusion coefficient. In other words, supercooled
water must show, on approaching $T_{g}$, the marked decoupling of
translational diffusion coefficient from viscosity or rotational
correlation time, as recently observed in some supercooled liquids
\cite{ediger,fujara,swallen}. Here we confirm by means of NMR data
the existence of a FSC in supercooled water, as proposed by the
QENS and MD studies \cite{chen2004,chen2005,stanley} and show, for
the first time, that the Stokes-Einstein relation (SER) breaks
down in different ways on both the fragile and strong sides of
FSC.

\begin{figure}[tbp]
\begin{center}
\includegraphics[width=8.5 cm]{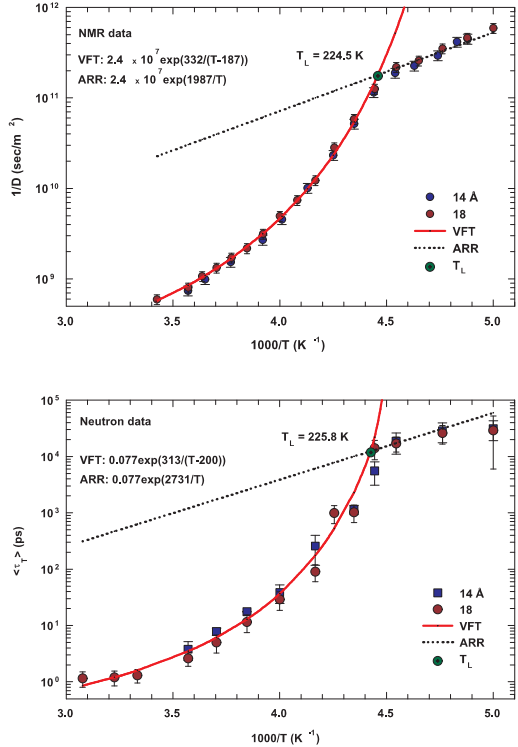}
\end{center}
\caption{Figure 1A shows, for the fully hydrated MCM-41-S samples
with diameter $14$ \AA\ and $18$ \AA, the inverse of the
self-diffusion coefficient of water $D$ measured by NMR as a
function of $1/T $ in a log-linear scale. The solid line denotes
the fit of the data to a Vogel-Fulcher-Tamman (VFT) relation. The
short dotted line denotes the fit to an Arrhenius law with the
same prefactor $1/D_{0}$. Figure 1B reports, as a function of
$1/T$, the average translational relaxation time $\left\langle
\tau _{T}\right\rangle $ obtained from QENS spectra in the same
experimental conditions of the NMR experiment. The dashed line
denotes the VFT law fit, and the dotted line the Arrhenius law fit
with the same prefactor $\tau _{0}$. In both panels, the values of
fitting parameters are shown.} \label{fig1}
\end{figure}

Figure 1A shows a log-linear plot of the inverse of the self-diffusion
coefficient of water $1/D$ measured by NMR as a function of $1/T$ for the
fully hydrated MCM-41-S samples with pore diameters of 14 \AA\ and 18 \AA .
Figure 1B reports the average translational relaxation time $\left\langle
\tau _{T}\right\rangle $, obtained by analyzing QENS spectra of the same
samples, versus $1/T$, using the Relaxing-Cage Model (RCM) as shown in Ref.
\cite{chen2004,chen2005}. As it can be observed from Figure 1A and 1B, the
measured values of $D$ and $\left\langle \tau _{T}\right\rangle $ are
independent of the pore size of the samples. This indicates that NMR
field-gradient measurements, having a length scale larger than the size of
the pores, are insensitive to the system geometry. In Fig. 1A, the solid
line denotes the fit of the data to a Vogel-Fulcher-Tamman (VFT) law $%
1/D=1/D_{0}exp(BT_{0}/(T-T_{0}))$, where $1/D_{0}=2.4\cdot 10^{7}$ ($s/m^{2}$%
), $B=1.775,$ and $T_{0}=187$ K. $B$ is a constant providing a measure of
the system fragility and $T_{0}$ the ideal glass transition temperature. The
short dotted line denotes the fit to an Arrhenius law $%
1/D=1/D_{0}exp(E_{A}/k_{B}T)$, where we keep the same $1/D_{0}$ value as in
the VFT law fit, and $E_{A}=3.98$ Kcal/mol. Figure 1B shows the $%
\left\langle \tau _{T}\right\rangle $ data at ambient pressure. The dashed
lines denote the VFT law fit, and the dotted lines the Arrhenius law fit,
with the same pre-factor $\tau _{0}$. We obtained the following values: $%
E_{A}=5.4$ Kcal/mol, $T_{0}=200$ K. The consequence of insisting on the same
pre-factor in both the VFT and the Arrhenius laws results in an equation
determining the crossover temperature $T_{L}$ in the following form: $%
1/T_{L}=1/T_{0}-Bk_{B}/E_{A}$. We obtained $T_{L}=224.5$ K from the $1/D$
data and $T_{L}=225.8$ K from the $\left\langle \tau _{T}\right\rangle $
data. The agreement between NMR and QENS results is thus satisfactory,
especially regarding the two relevant quantities $E_{A}$ and $T_{L}$. Figure
1A and 1B also show that within the size range of 10 to 20 \AA , the
crossover temperature $T_{L}$ is independent of pore size. As previously
mentioned, a FSC occurring at $228$ K has been proposed by Ito et al. \cite
{ito} for water, which is fragile at room and moderately supercooled
temperatures but near the glass transition temperature it is shown to be a
strong liquid by dielectric relaxation measurements \cite{bergman}. The
interpretation of this transition as a variant of the structural arrest
transition (as predicted by the ideal mode coupling theory) was the essence
of the recent QENS study of the structural relaxation time and MD study of
the self-diffusion coefficient \cite{chen2004,chen2005,stanley}. These NMR
results presented above constitute, by means of a direct measurement of the
self-diffusion coefficient of supercooled water, an independent confirmation
of the existence of FSC in water.

\begin{figure}[tbp]
\begin{center}
\includegraphics[width=8.5 cm]{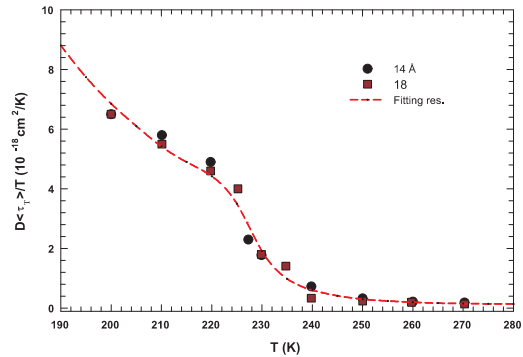}
\end{center}
\caption{The figure reports the quantity $D\left\langle \tau
_{T}\right\rangle /T$ as a function of $T$. Dots and squares
represent its values coming from the experimental data of $D$ and
$\left\langle \tau_{T}\right\rangle $ in samples with diameter
$14$ \AA\ and $18$ \AA, respectively. The dotted line represents
same quantity obtained by using the fitting values obtained from
the data reported in Fig 1A and 1B.} \label{fig2}
\end{figure}

Let us now focus on the SER that relates the self-diffusion coefficient $D$,
viscosity $\eta $, and temperature $T$ as $D\propto T/\eta $, which, as it
is well known, is usually accurate for normal and high temperature liquids.
Since $\left\langle \tau _{T}\right\rangle $ is proportional to the
viscosity, we examine the relationship between $D$ and $\left\langle \tau
_{T}\right\rangle $ in figure 2. In this figure the quantity $D\left\langle
\tau _{T}\right\rangle /T$ is reported as a function of $T$. Dots and
squares represent its values coming from the experimental data of samples
with pore diameters of $14$ and $18$ \AA , respectively, whereas the dotted
line represents the same quantity obtained using the corresponding fitted
lines reported in figure 1A and 1B. The temperature dependence of $%
D\left\langle \tau _{T}\right\rangle /T$ shows that this quantity is
constant at higher $T$, but increases steeply as $T$ goes below the FSC
temperature. Furthermore, it shows a small bump at the FSC temperature, in
accordance with the predictions of a recent theoretical study \cite{pan}.
Therefore, in the supercooled region the temperature behavior of $D$ and $%
\left\langle \tau _{T}\right\rangle $ is inconsistent with SER,
signalling a marked decoupling between these two transport
parameters, on decreasing $T$. In recent experimental studies on
some supercooled liquids, it has been reported that SER breaks
down as the glass transition is approached. The self-diffusion
coefficient shows, as far as water in the present experiment is
concerned, an enhancement of orders of magnitude from what
expected from SER \cite{ediger,fujara,swallen,chang,cicerone}.
These decouplings of the transport coefficients, observed as a SER
violation, have been attributed to the occurrence of dynamical
heterogeneities in structural glass formers \cite
{ediger,swallen,wolynes,ngai}. Thus, in supercooled liquids there
exist regions of varying dynamics, i.e. fluctuations that dominate
their transport properties near the glass transition. The extent
of such decouplings may depend on the material and the microscopic
details of the specific transport parameters.

\begin{figure}[tbp]
\begin{center}
\includegraphics[width=8.5 cm]{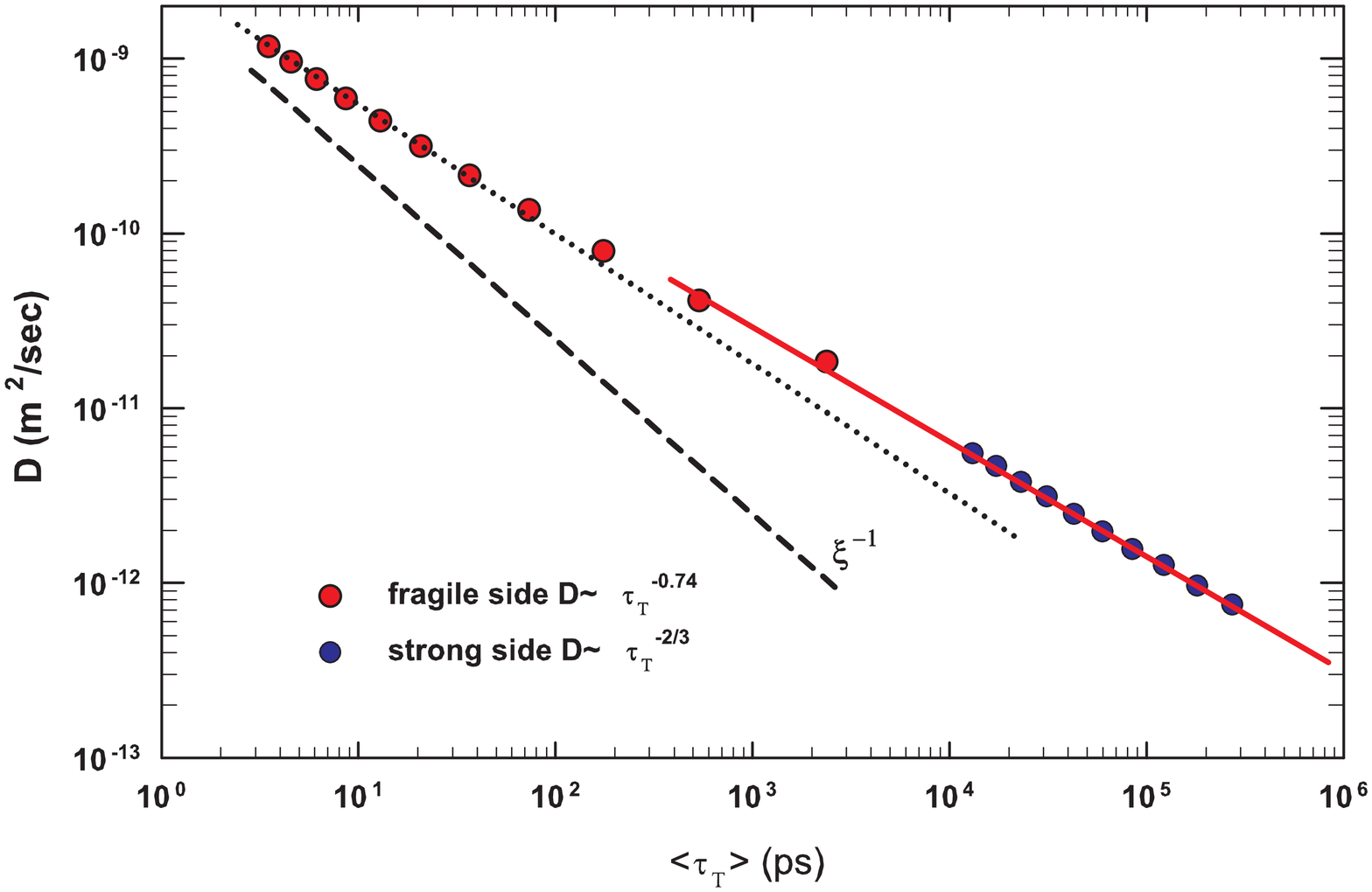}
\end{center}
\caption{The scaling plot in a log-log scale of $D$ vs $%
\left\langle \tau _{T}\right\rangle $. Red dots are data
corresponding to temperatures above $T_{L}$, i.e. when water is in
the fragile glass phase, whereas blues dots correspond to the
strong Arrhenius region. Two different scaling behaviors exist
above and below the temperature of the FST. In the
fragile region the scaling exponent is $\xi \simeq 0.74$ (dotted line) and $%
\sim 2/3$ in the strong side (solid line). Dashed line represent
the situation in which the SER holds, $D\sim \tau ^{-1}$.}
\label{fig3}
\end{figure}

The SER breakdown can be described using a scaling concept, in particular,
the law $D\sim \tau ^{-\xi }$, where $\xi =\alpha (T)/\beta (T)$ with $%
\alpha $ and $\beta $ being temperature dependent scaling exponents of $D$
and $\tau $, respectively \cite{jung}. Recently, it has been shown that for
tris-naphthylbenzene (a fragile glass former) $\xi =0.77$ \cite{swallen},
whereas an MD simulation of Lennard-Jones binary mixture has given $\xi =0.75
$ \cite{yamamoto}. By using such an approach, we will discuss our SER
results for confined supercooled water. Figure 3 shows the $D$ vs $%
\left\langle \tau _{T}\right\rangle $ plot in a log-log scale. The red dots
represent data corresponding to temperatures above $T_{L}$, where water
behaves as a fragile glass former, and the blue dots pertain to the strong
Arrhenius region. As it can be observed, the data clearly show two different
scaling behaviors above and below the FSC temperature, in particular $\xi
\simeq 0.74$ on the fragile side (dotted line) and $\sim 2/3$ on the strong
side (solid line). Dashed line represents the situation in which SER holds, $%
D\sim \tau ^{-1}$. These results agree with those obtained in
tris-naphthylbenzene \cite{swallen} and, more specifically, with those of a
recent theoretical study in which the decoupling of transport coefficients
in supercooled liquids was investigated by using two class of models, one
describing diffusion in a strong glass former, and the other in a fragile
one \cite{jung}. The main result of this study is that, while in the fragile
case the SER violation is weakly dependent on the dimensionality $d$, with $%
\xi =0.73$, in the strong case the violation is sensitive to $d$, going as $%
D\sim \tau ^{-2/3}$ for $d=1$, and as $D\sim \tau ^{-0.95}$ for $d=3$. On
considering the geometry of the system that we have used in our experiment
to confine water (1D cylindrical tubes, with a length of some $\mu m$ and
pore diameters of $14$ \AA\ and $18$ \AA ), the scaling plot shown in Figure
3 compares remarkably well with the findings of the theoretical
investigation \cite{jung}, on both fragile and strong sides.

In summary, we explore dynamical properties of water in a deeply supercooled
regime (well inside the ``no-man's land'') by means of NMR\ and QENS
experiments, which separately give a conclusive proof of the existence of a
FSC. This supports the hypothesis that liquid water is consisting of a
mixture of two different liquid structures (the LDL and HDL phases).
Accordingly, a liquid-liquid phase separation line exists in the $P-T$ plane
with a liquid-liquid critical point as its end point. Remarkably, we give
the first experimental proof of the existence of a violation of SER above
and below the FSC in water, i.e. in both the fragile and strong regimes of
supercooled water. This clearly reflects the decoupling of transport
coefficients of the liquid when temperature is lowered toward $T_{g}$. This
latter result certainly constitutes a new element which serves to clarify
one of the most intriguing properties of water.

\section*{Methods}

Water was confined in micellar templated mesoporous silica
matrices MCM-41-S, which have 1-D cylindrical pores with a length
of some $\mu m$ arranged in 2-D hexagonal arrays, synthesized
following a similar method for the previous synthesis of MCM-48-S
\cite{shih}. The MCM-41-S materials are the same as those
previously used in the QENS study of confined water \cite
{chen2005}. Pore size was determined using nitrogen
absorption-desorption technique \cite{chen2004,chen2005}.
Investigated samples have hydration levels of $H\simeq 0.5$ ($0.5$
gram H$_{2}$O per gram of MCM-41-S), obtained by exposing dry
powder samples to water vapor at room temperature in a closed
chamber. This water confining system can be regarded as one of the
most suitable adsorbent models currently available
\cite{schreiber,morishige}.

High-resolution QENS spectroscopy method was used to determine the
temperature dependence of $\left\langle \tau _{T}\right\rangle $ for
confined water. Because neutrons can easily penetrate the wall of sample
cell and because they are predominantly scattered by hydrogen atoms in
water, rather than by the matrices containing them, incoherent QENS is an
appropriate tool for our study. Using two separate high-resolution QENS
spectrometers, we were able to measure the translational-relaxation time
from 0.2~$ps$ to 10,000~$ps$ over the whole temperature range under study.
The experiments were performed at both the High-Flux Backscattering (HFBS)
and the Disc-Chopper Time-of-Flight (DCS) spectrometers in the NIST Center
for Neutron Research (NIST NCNR). The two spectrometers have two widely
different dynamic ranges (for the chosen experimental setup), one with an
energy resolution of 0.8~$\mu $eV (HFBS) and a dynamic range of $\pm $ 11~$%
\mu $eV~\cite{Meyer}, and the other with an energy resolution of 20~$\mu $eV
(DCS) and a dynamic range of $\pm $ 0.5~meV~\cite{JRC} in order to be able
to extract the broad range of relaxation times from the measured spectra.
The experiment was done at a series of temperatures, covering both below and
above the transition temperature, and the data were analyzed using
Relaxing-Cage Model to extract the average translational relaxation time $%
\left\langle \tau _{T}\right\rangle $.

The NMR experiments on fully hydrated MCM-41-S samples with pore
diameters of $18$ and $14$ \AA , were performed at ambient
pressure using a Bruker AVANCE NMR spectrometer, operating at
$700$ MHz proton resonance frequency. The self-diffusion
coefficient of water $D$ was measured with the pulsed gradient
spin-echo technique (PGSE) \cite{tanner,Price1998}, in the
temperature range $190$ K - $298$ K (with an accuracy of $\pm $
$0.2$ K). The $T$ dependence of the chemical shift of methanol was
used as a $T$ standard. All details about the NMR experiment and
the sample properties are reported elsewhere \cite{malla}. The
reported $D$ values were derived from the measured mean square
displacement $\left\langle r^{2}(t)\right\rangle $ of molecules
diffusing along the NMR pulsed field gradients direction
$\mathbf{r}$, during the time interval $t$.

\section*{Acknowledgements}

Authors would like to thank Chun-Wan Yen from National Taiwan
University for preparing MCM-41-S samples. Technical supports in
QENS measurements from E. Mamontov and J.R.D. Copley at NIST NCNR
are greatly appreciated. The research at MIT is supported by DOE
Grants DE-FG02-90ER45429 and 2113-MIT-DOE-591. The research in
Messina is supported by the MURST-PRIN2004. This work utilized
facilities supported in part by the National Science Foundation
under Agreement No. DMR-0086210. The work utilized facilities of
the Messina SCM-HR-NMR Center of CNR-INFM. We benefited from
affiliation with EU Marie-Curie Research and Training Network on
Arrested Matter.

\end{document}